\documentclass[
]{ceurart}

\usepackage{listings}
\usepackage{booktabs}
\usepackage{xcolor}

\usepackage{anyfontsize}
\usepackage{soul}
\begin{document}

\copyrightyear{2022}
\copyrightclause{Copyright for this paper by its authors.
  Use permitted under Creative Commons License Attribution 4.0
  International (CC BY 4.0).}

\conference{eCom’24: ACM SIGIR Workshop on eCommerce, July 18, 2024, Washington, D.C., USA}

\title{Doc2Token: Bridging Vocabulary Gap by Predicting Missing Tokens for E-commerce Search}


\author[1]{Kaihao Li}[%
email=kaihao.li@walmart.com
]
\address[1]{Walmart Global Technology, USA}

\author[1]{Juexin Lin}[%
email=juexin.lin@walmart.com
]

\author[1]{Tony Lee}[%
email=tony.lee@walmart.com
]

\begin{abstract}
  Addressing the ``vocabulary mismatch'' issue in information retrieval is a central challenge for e-commerce search engines, because product pages often miss important keywords that customers search for. Doc2Query~\cite{nogueira2019document} is a popular document-expansion technique that predicts search queries for a document and includes the predicted queries with the document for retrieval. However, this approach can be inefficient for e-commerce search, because the predicted query tokens are often already present in the document. In this paper, we propose Doc2Token, a technique that predicts relevant tokens (instead of queries) that are missing from the document and includes these tokens in the document for retrieval. For the task of predicting missing tokens, we introduce a new metric, ``novel ROUGE score''. Doc2Token is demonstrated to be superior to Doc2Query in terms of novel ROUGE score and diversity of predictions. Doc2Token also exhibits efficiency gains by reducing both training and inference times. We deployed the feature to production and observed significant revenue gain in an online A/B test, and launched the feature to full traffic on Walmart.com.
\end{abstract}


\maketitle

\section{Introduction}
\label{sec:intro}

The vocabulary gap problem in e-commerce search is a central challenge, as it arises from discrepancies between the vocabulary used by customers and sellers when describing products. Customer queries are often short and ambiguous, while product descriptions tend to be more detailed and explicit. For instance, a customer might search for ``small building set'', intending to find a set that offers simpler building experiences for young children. However, in the product catalog, those products are often characterized by piece count and target age group, which do not align directly with this search query. 

Different approaches have been proposed to address the vocabulary mismatch issue. In the context of lexical retrieval, query expansion~\cite{rocchio71relevance, miller1995wordnet, lv2009comparative, liu2022query, wang2023query2doc} and document expansion~\cite{nogueira2019document, nogueira2019doc2query, formal2021splade} are two effective techniques. Query expansion enriches user queries with additional terms or synonyms to better capture the user's intent, while document expansion enriches product information with additional keywords or phrases. Doc2Query~\cite{nogueira2019doc2query} is a document expansion technique that predicts and indexes search queries for documents. Recently, embedding-based dense retrieval models~\cite{xiong2020approximate} have demonstrated their ability to align queries and documents by projecting them into a representation space to learn their semantic similarity. Although dense retrieval with approximate nearest neighbor search has shown impressive results, lexical retrieval remains an important component of e-commerce search due to its desirable properties, such as interpretability, scalability, as well as handling of rare words and numerical tokens. 

\begin{figure}[t] 
\centering
\includegraphics[width=0.8\textwidth,height=0.32\textheight]{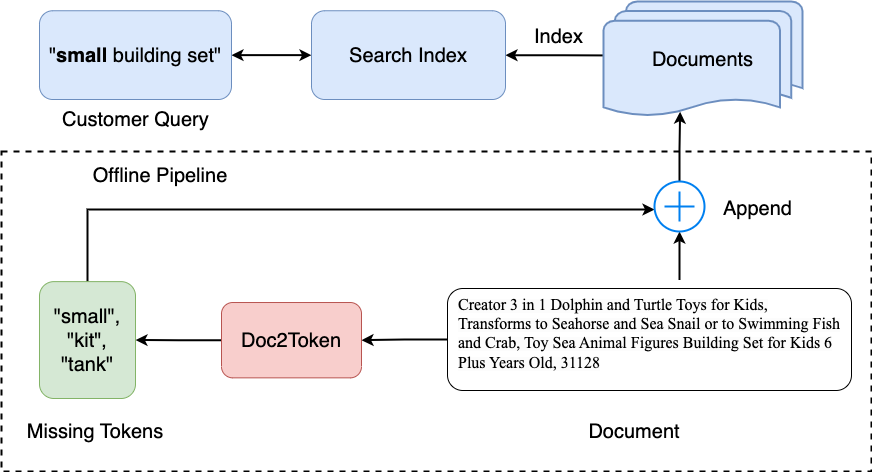} 
\caption{Doc2Token Overview.}
\label{overview}
\vspace{-2mm}
\end{figure} 

In this paper, we propose a new document expansion technique, called Doc2Token, which generalizes Doc2Query for application in e-commerce search, as depicted in Figure~\ref{overview}.  
Our task is to, given a product, generate relevant keywords that are absent from the product's indexed metadata to ensure that the product is retrieved when customers search using these keywords. We call these ``novel tokens''. We observed that Doc2Query's predicted queries often contain tokens already in the product metadata instead of novel tokens, which makes it inefficient for our task. In contrast, Doc2Token is designed to predict novel tokens (instead of queries). The approach is to prepare a dataset of pairs of product and novel token, then train a seq2seq model on that dataset. By design, Doc2Token efficiently generates tokens with a high probability of being novel, rather than producing long and redundant sequences like in Doc2Query. Using the product shown in Figure~\ref{overview} as an example, Doc2Query predicts ``6 year old boy toy'', ``3 in 1 creator'', ``sea animal toy'', and ``building toy for boy''. On the other hand, Doc2Token provides a more diversified set of tokens, including ``small'', ``kit'', and ``tank''. To incorporate the Doc2Token predictions into the search system, we added them to the product's metadata and indexed for retrieval matching and ranking. 

To assess performance on the novel-token-prediction task, we introduce a new evaluation metric called ``novel ROUGE score'', denoted by ``nROUGE'', to measure the ROUGE score~\cite{lin2004rouge} specifically for novel tokens. Our results indicate that Doc2Token surpasses Doc2Query in terms of nROUGE score. Regarding efficiency, Doc2Token is capable of predicting more diverse results while significantly reducing training and inference times. The effectiveness of this approach is further demonstrated through online relevance evaluation and A/B testing, confirming that the novel tokens generated are not only novel but also relevant to products, ultimately driving customer engagement. 

Our contributions are summarized as follows:
\begin{itemize}
    \item We propose a novel technique, Doc2Token, for document expansion in e-commerce search, encompassing both the training setup and the modification of the loss function.
    \item We introduce a new metric, ``novel ROUGE score'', to evaluate the performance of predicting novel tokens.
    \item We demonstrate that Doc2Token achieves improvements in both effectiveness and efficiency compared to Doc2Query.
\end{itemize}
\section{Methodology}
\label{sec:methods}
We define the task of novel token prediction in this section. For a product $i$, let $p_i$ and $T_i$ represent the product text and the set of tokens extracted from $p_i$, respectively. The goal is to predict novel tokens, i.e., tokens that are absent from $T_i$. To achieve this, we first collect a list of relevant queries (section~\ref{subsec:data}) for the product based on historical search logs. Next, for each product, we assemble a set of unique tokens, disregarding their sequence, through the subsequent process. We concatenate all queries, divide the concatenated sequence into individual tokens, count their frequencies, and exclude tokens already present in product $i$.  As a result, for product $i$, we have a target token set $\bigcup_{t_{ik} \notin T_{i}}{(t_{ik}, f_{ik})}$, where $t_{ik}$ represents the $k$th unique token from target queries for product $i$, and $f_{ik}$ denotes the frequency of token $t_{ik}$. Instead of using all tokens as one training target, we divide them into $k$ training instances. 
We train a seq2seq generative language model, T5~\cite{raffel2020exploring}, with an encoder-decoder structure. It takes the product text $p_i$ as input and outputs a sequence of tokens $\hat{t}_i$ in an autoregressive manner. More formally, 
\begin{equation} \label{eq.t5}
\hat{t}_i = \mbox{Decoder} ( \mbox{Encoder}(p_{i})).
\end{equation}
To account for the token frequency, we modify the T5 loss as follows. The loss for a product-token pair is:
\begin{equation} \label{eq.loss}
loss_{\mbox{weighted}}(t_{ik}, \hat{t}_i) = (f_{ik})^\alpha * loss_{T5}(t_{ik}, \hat{t}_i),
\end{equation}
where $\alpha$ is the smoothing factor, set to $0.5$ in our implementation. This value is chosen to balance the contribution of token frequency to the overall loss. In practice, we use the cross-entropy loss for $loss_{T5}$.

For model inference, we employ beam search \cite{yang2018breaking} to generate the top N predictions. 
Since T5 tokenizes words into subtokens \cite{kudo2018sentencepiece}, the target token $t_{ik}$ and predicted token $\hat{t}_i$ may consist of multiple subtokens, although they are always single words. We utilize the beam score \cite{yang2018breaking} to determine the confidence of the prediction, and we only retain predictions with scores greater than a predetermined cutoff value (more details are discussed in Section~\ref{subsec:results}).

\section{Experiments}

\subsection{Datasets}
\label{subsec:data}
We sampled the product-query pairs from user engagement data on Walmart.com with at least a certain number of add-to-cart (ATCs) over a two-year period. Then we did the following preprocessing steps. The product information used throughout the experiment includes product title, product type, brand, color, gender and description.

\noindent \textit{Relevance filter (RF).} The engaged products are not always relevant to the search query, because users’ decisions are influenced by factors other than relevance, such as price, visual appeal, ranking, etc. Additionally, the minimal match criteria~\cite{shahi2016apache,lucene} for our lexical retrieval is not always 100\%. As a result, customers may be shown products that don't fully match their search terms. For instance, a customer may search for ``vanilla ice cream'' but end up buying chocolate ice cream. To mitigate such noise, we removed product-query pairs predicted to be irrelevant by applying a relevance model. (The relevance model is a BERT~\cite{bert} cross-encoder model with a classifier head that takes the query and product information as input and outputs a relevance score. It was trained on manually annotated relevance data.) 

\noindent \textit{Full match filter (FMF).} 
For the purpose of predicting novel tokens, we focused on product-query pairs with a vocabulary mismatch. This implies that at least one query token was not found in the product information. Thus, we removed pairs where all query tokens were in the product information. 

\noindent \textit{Price token filter (PTF).} Customer queries sometimes include price and deal intent (e.g., ``under \$500'' or ``on sale''). Such phrases are not very useful for our task, since price and deals can fluctuate rapidly, so it does not make sense to include them as training labels. We utilized regular expressions to identify and eliminate these phrases from the query. 

\noindent \textit{Overlapping token filter (OTF).} This step excludes all query tokens that are present in the product information. (This is a stronger extension of the full match filter.) After this step, only novel tokens remain.

Detailed data statistics can be found in Table~\ref{tbl:stat}. In later sections, we show results for both Doc2Query and Doc2Token. For the Doc2Query dataset, we applied the first three filters, resulting in 14.9M product-query pairs. For the Doc2Token dataset, we built upon the Doc2Query dataset by further dividing the queries into tokens and applying the overlapping token filter, resulting in 10.3M product-token pairs. For each dataset, we partitioned it by product into training, validation, and test sets in an 8:1:1 ratio to ensure no product overlap between the sets.

\begin{table}[t]
\small
\begin{center}
\begin{tabular}{p{5.5cm}|p{3cm}|p{3cm}|p{1.3cm}} 
\hline
Preprocessing & product-query pairs & product-token pairs & products \\
\hline
\hline
None & 81.8M & & 8.7M   \\
\hline
RF & 66.4M & & 8.0M \\
\hline
RF + FMF + PTF \newline (Doc2Query) & 14.9M & & 3.1M  \\
\hline
RF + FMF + PTF + tokenization + OTF \newline (Doc2Token) & & 10.3M & 3.1M \\
\hline
\end{tabular}
\caption{Summary statistics for each data preprocessing step.}\label{tbl:stat}
\end{center}
\vspace{-3mm}
\end{table}


\subsection{Metrics}
\label{subsec:metrics}
To evaluate model performance, we utilize the standard ROUGE score, a widely-used evaluation metric for summarization tasks \cite{see2017get}. The ROUGE score assesses the quality of generated text by comparing it to a reference text, considered as ground truth, using n-gram overlaps. Unlike the summarization task, which may require complex evaluations based on n-grams, lexical retrieval primarily relies on unigrams. Therefore, we measure the ROUGE scores in terms of unigrams. In our context, we formulate the ROUGE score as follows.
\begin{equation}
    ROUGE_{Precision} = \frac{1}{N} \sum_i \frac{\sum_{token \in y_i} count_{match}(token, \hat{y}_i)}{count(\hat{y}_i)},
\end{equation}
\begin{equation}
    ROUGE_{recall} = \frac{1}{N} \sum_i 
 \frac{\sum_{token \in y_i} count_{match}(token, \hat{y}_i)}{count(y_i)},
\end{equation}
where $y_i$, $\hat{y}_i$ represent the reference text and predicted text for product $i$, respectively. $count$ denotes the number of tokens, and $count_{match}$ denotes the number of co-occurrences of a reference token in the predicted text. $N$ is the number of products.

However, a higher ROUGE score does not necessarily indicate better performance at predicting novel tokens. We observed that text predictions with high ROUGE scores often exhibit substantial token overlap with the product information. To assess the performance for novel tokens, we introduce a new metric, ``novel ROUGE score'' (nROUGE), where the reference text consists solely of novel tokens. Formally, 
the nROUGE score is defined as follows. 
\begin{equation}
    nROUGE_{precision} = \frac{1}{N} \sum_i \frac{\sum_{token \in y^*_i} count_{match}(token, \hat{y}_i)}{count(\hat{y}_i)},
\end{equation}
\begin{equation}
    nROUGE_{recall} = \frac{1}{N} \sum_i \frac{\sum_{token \in y^*_i} count_{match}(token, \hat{y}_i)}{count(y^*_i)},
\end{equation}
where $y^*_i$ represents the novel reference text for product $i$, i.e., the tokens in the reference text but not in the product information.

\subsection{Experiment Setup}
For the model training, we fine-tuned the public T5-base model using eight A-100 Nvidia GPUs, learning rate of 1e-4, batch size of 64, maximum input sequence of 256, and maximum output sequence of 32. For model inference, we employed a top-k beam search strategy with a beam size of 10. The input of the model is a text of product information consisting of the product title and some attributes, such as brand, color, etc.

\subsection{Results}
\label{subsec:results}
In this section, we compare our proposed Doc2Token model to the baseline Doc2Query model from both effectiveness and efficiency perspectives. To assess effectiveness, we measure performance based on the ROUGE and nROUGE scores. In terms of efficiency, we report the resources used for model training and inference. We evaluated on four models, including both Doc2Query and Doc2Token models with and without full-match filter in data preprocessing step.

\subsubsection{Offline evaluation results}
In Table~\ref{tbl.rouge}, we present the results for each model based on the top 10 predictions with various beam score cutoffs. The predictions were chosen if their beam scores exceeded the respective cutoff value. These cutoff values were tuned, for each model, to achieve the optimal nROUGE F1 score. Additionally, to assess models' efficiency in predicting novel tokens, we concatenated the predictions and calculated the total number of predicted tokens and the number of predicted novel tokens. For the Doc2Query models, we present the result without any beam score cutoff, as well as the result that achieves the highest nROUGE F1 score. For the Doc2Token models, we reported three results: one without any cutoff, one with the optimal nROUGE F1 score, and a result generating a similar number of novel tokens as the optimal Doc2Query model.

\begin{table*}
\small
    \centering
    \begin{tabular}{c|c|ccc|ccc|rrr}
    \hline
        \multirow{2}{*}{Model} & \multirow{2}{*}{Cutoff} & \multicolumn{3}{c}{ROUGE} & \multicolumn{3}{|c}{nROUGE} & \multicolumn{3}{|c}{\# of predicted tokens}\\ 
        \cline{3-11}
        & & Precision &Recall &F1& Precision & Recall &F1&Total&Novel& \% \\
        \hline
        \hline
        Doc2Query & 0.00 & 0.505 & 0.704 & 0.533 & 0.409 & 0.606 & 0.427
        & 28.9 & 9.0 & 31 \\ 
        w/o FMF & 0.34 & 0.535 & 0.680 & 0.545 & 0.434 & 0.574 & 0.438
        & 25.5 & 8.3 & 33 \\ \hline
        Doc2Token & 0.00 & 0.448 & 0.696 & 0.596 & 0.346 & 0.636 & 0.393
        & 10.0 & 10.0 & 100 \\ 
         w/o (FMF & 0.15 & 0.534 & 0.664 & 0.548 & 0.453 & 0.593 & 0.464 
        & 8.2 & 8.2 & 100  \\
        + OTF) & 0.18 & 0.564 & 0.640 & 0.556 & 0.479 & 0.560 & 0.469 
        & 7.5 & 7.5 & 100  \\ 
        \hline 
        
        Doc2Query & 0.00 & 0.418 & 0.797 & 0.496 & 0.338 & 0.755 & 0.402 
        & 31.8 & 6.2 & 19 \\ 
        w/ FMF &  0.51 & 0.558 & 0.673 & 0.554 & 0.497 & 0.622 & 0.481 
        & 15.2 & 3.2 & 21 \\ 
        \hline
        
        Doc2Token  & 0.00 & 0.214 & 0.364 & 0.240 & 0.205 & 0.767 & 0.286
        & 10.0 & 9.8 & 98  \\
        w/ (FMF & 0.29 & 0.517 & 0.281 & 0.329 & 0.507 & 0.623 & 0.496 
        & 3.2 & 3.2 & 100  \\ 
        + OTF) & 0.33 & 0.553 & 0.261 & 0.321 & 0.554 & 0.583 & \textbf{0.500} 
        & 2.6 & 2.6 & 100  \\
        \hline
        
    \end{tabular}
    \caption{Results to compare Doc2Query models and Doc2Token models. }\label{tbl.rouge}
\end{table*}

To evaluate the effectiveness of the full-match filter in data preprocessing, we trained the models without incorporating that step. We observed no apparent impact on the ROUGE F1 score. However, there was a substantial improvement in the nROUGE F1 score for both Doc2Query (from 0.438 to 0.481) and Doc2Token (from 0.469 to 0.500) at the optimal cutoff values. This is expected, as with the full-match filter, our training data primarily relies on labels containing novel tokens.

In our comparison between the Doc2Query and Doc2Token models, we observed that the Doc2Query models tend to achieve higher ROUGE F1 scores than the Doc2Token models. However, Doc2Token excels in achieving superior nROUGE F1 scores. 
Comparing the models with a similar number of predicted novel tokens, for example, the Doc2Query model with a cutoff of 0.51 and Doc2Token model with a cutoff of 0.29, the Doc2Token model outperforms the Doc2Query model in both nROUGE precision and nROUGE recall, yielding a higher nROUGE F1 score.
With optimal cutoffs, the Doc2Token model shows the superior performance compared to the Doc2Query model by achieving a 3.95\% higher nROUGE F1 score from 0.496 to 0.500. This improvement is statistically significant, with a 95\% confidence interval for the Doc2Token F1 score of (0.498, 0.501) obtained through bootstrap resampling.
Moreover, the Doc2Token model is more efficient in generating novel tokens, achieving nearly 100\% of predicted tokens being novel. In contrast, Doc2Query produces only 20\% novel tokens, indicating a higher degree of redundancy. This is expected, as the Doc2Token model is designed to predict more diverse novel tokens.

\subsubsection{Model efficiency}
\label{sec:time}
\begin{table}[t]
\small
	\centering
	\vspace{-3mm}
	\begin{tabular}{c|c|cc}
	\hline
	Model & Preprocessing &\makecell{Train Time} & Inference Time \\
	\hline
        \hline
        Doc2Query & w/o FMF & 1002 & 136 \\
        \hline
        Doc2Token & w/o (FMF + OTF) & 1416 & 76 \\
        \hline
        Doc2Query & all & 166 & 141 \\
        \hline
	Doc2Token & all & \textbf{108} & \textbf{76} \\
	\hline
	\end{tabular}
        \caption{Results to compare model train and inference time. Train time and inference time are measured at minutes per one epoch and minutes per 100k products, respectively.} \label{tbl.efficiency}
\vspace{-3mm}
\end{table}

Table ~\ref{tbl.efficiency} presents the results for training and inference times. The training time is primarily affected by the size of the training data. Without the full-match filter, splitting queries into tokens explodes the data size, resulting in a longer training time for Doc2Token compared to Doc2Query. However, with the full-match filter, the situation changes: the Doc2Token strategy significantly reduces the dataset size, leading to shorter training times than Doc2Query. For inference time, we sampled 100,000 products from the test dataset and conducted model inference on the top 10 results with a batch size of 16 using a single K80 GPU machine. The inference time for Doc2Token is faster than that for Doc2Query, as the output of Doc2Token is generally shorter. The results are in agreement with the efficiency discussions from Table~\ref{tbl.rouge}.

\subsection{Examples}
\begin{table*}[ht]
\small
    \centering
    \begin{tabular}{p{0.8in}|p{4.6in}}
    \hline
        Product Input & title: Toddler Floaties, Swim Vest for Boys and Girls Age 2-7 Years Old, 20-50 Pounds Children Water Wings Arm Floaties in Puddle/Sea/Pool/Beach (Dinosaur)
        \newline brand: Dark Lightning color: Blue gender: Unisex \\ \hline
        Doc2Query & ``swimming vest for \textbf{kid}'', ``toddler boy swim vest'', ``swim vest for \textbf{kid}'', ``boy floaty'', ``\textbf{kid} floaty'' \\ \hline
        Doc2Token & ``\textbf{float}'', ``\textbf{kid}'', ``floaty'', ``\textbf{floater}'', ``\textbf{salvavida}'', ``\textbf{swimmy}'', ``\textbf{baby}'', ``\textbf{floatation}'', ``children'', ``\textbf{life}'' \\ \hline
        \hline 
        Product Input & title: Hanno Muller-Brachmann - North German Poets - Classical - CD
        \newline brand: Artists  color: white \\ \hline
        Doc2Query & ``classical cds'', ``germany cd'', ``\textbf{country} music cd'', ``\textbf{north} germany cd'', ``\textbf{west} germany cd'' \\ \hline
        Doc2Token & ``\textbf{music}'', ``\textbf{country}'', ``\textbf{5}'', ``\textbf{b}'', ``classical'', ``\textbf{christmas}'', ``\textbf{soundtrack}'' \\ \hline
    \end{tabular}
    \caption{Two product examples for Doc2Query and Doc2Token: A good example \& A case for improvement}\label{tbl.example}
\end{table*}

Table~\ref{tbl.example} showcases two example products along with its corresponding Doc2Query and Doc2Token predictions. The Doc2Query model produces the top 5 queries, while the Doc2Token model generates the top 10 tokens. The novel tokens, after the process of stemming ~\cite{shahi2016apache,lucene}, are bold. In general, the Doc2Query model produces queries containing tokens that are already present in the product information, while the Doc2Token model does not. In contrast, all tokens produced by the Doc2Token model are relevant and absent from the product. 
In the first positive example, the Doc2Token model is capable of predicting a Spanish word ``salvavida'' (``lifeguard'' in English), indicating its ability to handle Spanish queries. Queries in Spanish are commonly observed in US e-commerce search. The second example, a product from northern Germany, illustrates some bad predictions from the models. The predicted tokens ``country'', ``west'', ``christmas'' are irrelevant. This is mainly due to a lack of media-related data in our training set. Replacing T5 with a more knowledgeable LLM could potentially address this issue. 

\section{Implementation and online tests}

We implemented the Doc2Token model in production because of its superior performance compared to the Doc2Query model as shown in Section~\ref{subsec:results}. We ran the model inference on all products in our catalog using cost-effective K80 GPUs and a batch size of 16. We predicted the top 10 tokens and retained the predictions with scores above 0.33. 
The inference process is conducted offline on a daily basis. For online usage, the Doc2Token predictions serve as an additional text matching field in Solr~\cite{shahi2016apache}, which is an enterprise search platform built on Apache Lucene~\cite{lucene} utilized for the search retrieval at Walmart.com.

We evaluated the performance of the Doc2Token feature from both relevance and engagement perspectives. For the relevance evaluation, we enlisted the human annotators to assess the top 10 ranked products from impacted queries. These assessments were based on a three-point scale (exact match, substitute, irrelevant), considering factors such as product title, image, and product page at Walmart.com. We then computed the NDCG@10 based on this 3-point scale, showing a 0.49\% lift (p-value=0.066). For engagement assessment, we conducted a two-week A/B test for the feature on live traffic. The test revealed a statistically significant 0.28\% lift in revenue (p-value = 0.013). While the NDCG@10 improvement is statistically marginal, the statistically significant revenue increase demonstrates the effectiveness of the Doc2Token feature. By introducing relevant products in the retrieval process, the Doc2Token feature is able to enhance the end-to-end search results, assisting customers in finding what they are searching for.


    \section{Conclusions}
In this study, we present Doc2Token, a novel document expansion technique for e-commerce search engines. We introduce the novel ROUGE score, a new metric crafted to evaluate the efficacy of document expansion endeavors. Our analysis has demonstrated that Doc2Token surpasses Doc2Query in terms of efficiency and effectiveness in addressing the vocabulary mismatch challenge. 
The Doc2Token feature has been deployed and evaluated online, resulting in a significant improvement in both relevance and revenue.


\bibliography{paper}

\begin{thebibliography}{17}
\expandafter\ifx\csname natexlab\endcsname\relax\def\natexlab#1{#1}\fi
\providecommand{\url}[1]{\texttt{#1}}
\providecommand{\href}[2]{#2}
\providecommand{\path}[1]{#1}
\providecommand{\DOIprefix}{doi:}
\providecommand{\ArXivprefix}{arXiv:}
\providecommand{\URLprefix}{URL: }
\providecommand{\Pubmedprefix}{pmid:}
\providecommand{\doi}[1]{\href{http://dx.doi.org/#1}{\path{#1}}}
\providecommand{\Pubmed}[1]{\href{pmid:#1}{\path{#1}}}
\providecommand{\bibinfo}[2]{#2}
\ifx\xfnm\relax \def\xfnm[#1]{\unskip,\space#1}\fi
\bibitem[{Nogueira et~al.(2019)Nogueira, Yang, Lin, and Cho}]{nogueira2019document}
\bibinfo{author}{R.~Nogueira}, \bibinfo{author}{W.~Yang}, \bibinfo{author}{J.~Lin}, \bibinfo{author}{K.~Cho},
\newblock \bibinfo{title}{Document expansion by query prediction},
\newblock \bibinfo{journal}{arXiv preprint arXiv:1904.08375}  (\bibinfo{year}{2019}).
\bibitem[{Rocchio(1971)}]{rocchio71relevance}
\bibinfo{author}{J.~J. Rocchio},
\newblock \bibinfo{title}{Relevance feedback in information retrieval},
\newblock in: \bibinfo{editor}{G.~Salton} (Ed.), \bibinfo{booktitle}{The Smart retrieval system - experiments in automatic document processing}, \bibinfo{publisher}{Englewood Cliffs, NJ: Prentice-Hall}, \bibinfo{year}{1971}, pp. \bibinfo{pages}{313--323}.
\bibitem[{Miller(1995)}]{miller1995wordnet}
\bibinfo{author}{G.~A. Miller},
\newblock \bibinfo{title}{Wordnet: a lexical database for english},
\newblock \bibinfo{journal}{Communications of the ACM} \bibinfo{volume}{38} (\bibinfo{year}{1995}) \bibinfo{pages}{39--41}.
\bibitem[{Lv and Zhai(2009)}]{lv2009comparative}
\bibinfo{author}{Y.~Lv}, \bibinfo{author}{C.~Zhai},
\newblock \bibinfo{title}{A comparative study of methods for estimating query language models with pseudo feedback},
\newblock in: \bibinfo{booktitle}{Proceedings of the 18th ACM conference on Information and knowledge management}, \bibinfo{year}{2009}, pp. \bibinfo{pages}{1895--1898}.
\bibitem[{Liu et~al.(2022)Liu, Li, Lin, Riedel, and Stenetorp}]{liu2022query}
\bibinfo{author}{L.~Liu}, \bibinfo{author}{M.~Li}, \bibinfo{author}{J.~Lin}, \bibinfo{author}{S.~Riedel}, \bibinfo{author}{P.~Stenetorp},
\newblock \bibinfo{title}{Query expansion using contextual clue sampling with language models},
\newblock \bibinfo{journal}{arXiv preprint arXiv:2210.07093}  (\bibinfo{year}{2022}).
\bibitem[{Wang et~al.(2023)Wang, Yang, and Wei}]{wang2023query2doc}
\bibinfo{author}{L.~Wang}, \bibinfo{author}{N.~Yang}, \bibinfo{author}{F.~Wei},
\newblock \bibinfo{title}{Query2doc: Query expansion with large language models},
\newblock \bibinfo{journal}{arXiv preprint arXiv:2303.07678}  (\bibinfo{year}{2023}).
\bibitem[{Nogueira et~al.(2019)Nogueira, Lin, and Epistemic}]{nogueira2019doc2query}
\bibinfo{author}{R.~Nogueira}, \bibinfo{author}{J.~Lin}, \bibinfo{author}{A.~Epistemic},
\newblock \bibinfo{title}{From doc2query to doctttttquery},
\newblock \bibinfo{journal}{Online preprint} \bibinfo{volume}{6} (\bibinfo{year}{2019}) \bibinfo{pages}{2}.
\bibitem[{Formal et~al.(2021)Formal, Piwowarski, and Clinchant}]{formal2021splade}
\bibinfo{author}{T.~Formal}, \bibinfo{author}{B.~Piwowarski}, \bibinfo{author}{S.~Clinchant},
\newblock \bibinfo{title}{Splade: Sparse lexical and expansion model for first stage ranking},
\newblock in: \bibinfo{booktitle}{Proceedings of the 44th International ACM SIGIR Conference on Research and Development in Information Retrieval}, \bibinfo{year}{2021}, pp. \bibinfo{pages}{2288--2292}.
\bibitem[{Xiong et~al.(2020)Xiong, Xiong, Li, Tang, Liu, Bennett, Ahmed, and Overwijk}]{xiong2020approximate}
\bibinfo{author}{L.~Xiong}, \bibinfo{author}{C.~Xiong}, \bibinfo{author}{Y.~Li}, \bibinfo{author}{K.-F. Tang}, \bibinfo{author}{J.~Liu}, \bibinfo{author}{P.~Bennett}, \bibinfo{author}{J.~Ahmed}, \bibinfo{author}{A.~Overwijk},
\newblock \bibinfo{title}{Approximate nearest neighbor negative contrastive learning for dense text retrieval},
\newblock \bibinfo{journal}{arXiv preprint arXiv:2007.00808}  (\bibinfo{year}{2020}).
\bibitem[{Lin(2004)}]{lin2004rouge}
\bibinfo{author}{C.-Y. Lin},
\newblock \bibinfo{title}{Rouge: A package for automatic evaluation of summaries},
\newblock in: \bibinfo{booktitle}{Text summarization branches out}, \bibinfo{year}{2004}, pp. \bibinfo{pages}{74--81}.
\bibitem[{Raffel et~al.(2020)Raffel, Shazeer, Roberts, Lee, Narang, Matena, Zhou, Li, and Liu}]{raffel2020exploring}
\bibinfo{author}{C.~Raffel}, \bibinfo{author}{N.~Shazeer}, \bibinfo{author}{A.~Roberts}, \bibinfo{author}{K.~Lee}, \bibinfo{author}{S.~Narang}, \bibinfo{author}{M.~Matena}, \bibinfo{author}{Y.~Zhou}, \bibinfo{author}{W.~Li}, \bibinfo{author}{P.~J. Liu},
\newblock \bibinfo{title}{Exploring the limits of transfer learning with a unified text-to-text transformer},
\newblock \bibinfo{journal}{The Journal of Machine Learning Research} \bibinfo{volume}{21} (\bibinfo{year}{2020}) \bibinfo{pages}{5485--5551}.
\bibitem[{Yang et~al.(2018)Yang, Huang, and Ma}]{yang2018breaking}
\bibinfo{author}{Y.~Yang}, \bibinfo{author}{L.~Huang}, \bibinfo{author}{M.~Ma},
\newblock \bibinfo{title}{Breaking the beam search curse: A study of (re-) scoring methods and stopping criteria for neural machine translation},
\newblock \bibinfo{journal}{arXiv preprint arXiv:1808.09582}  (\bibinfo{year}{2018}).
\bibitem[{Kudo and Richardson(2018)}]{kudo2018sentencepiece}
\bibinfo{author}{T.~Kudo}, \bibinfo{author}{J.~Richardson},
\newblock \bibinfo{title}{Sentencepiece: A simple and language independent subword tokenizer and detokenizer for neural text processing},
\newblock \bibinfo{journal}{arXiv preprint arXiv:1808.06226}  (\bibinfo{year}{2018}).
\bibitem[{Shahi(2016)}]{shahi2016apache}
\bibinfo{author}{D.~Shahi}, \bibinfo{title}{Apache solr}, \bibinfo{publisher}{Springer}, \bibinfo{year}{2016}.
\bibitem[{luc(2019)}]{lucene}
\bibinfo{title}{Apache lucene}, \bibinfo{howpublished}{\url{http://lucene.apache.org}}, \bibinfo{year}{2019}.
\bibitem[{Devlin et~al.(2018)Devlin, Chang, Lee, and Toutanova}]{bert}
\bibinfo{author}{J.~Devlin}, \bibinfo{author}{M.~Chang}, \bibinfo{author}{K.~Lee}, \bibinfo{author}{K.~Toutanova},
\newblock \bibinfo{title}{{BERT:} pre-training of deep bidirectional transformers for language understanding},
\newblock \bibinfo{journal}{CoRR} \bibinfo{volume}{abs/1810.04805} (\bibinfo{year}{2018}).
\bibitem[{See et~al.(2017)See, Liu, and Manning}]{see2017get}
\bibinfo{author}{A.~See}, \bibinfo{author}{P.~J. Liu}, \bibinfo{author}{C.~D. Manning},
\newblock \bibinfo{title}{Get to the point: Summarization with pointer-generator networks},
\newblock \bibinfo{journal}{arXiv preprint arXiv:1704.04368}  (\bibinfo{year}{2017}).

\end{thebibliography}




\end{document}